# REMOTE SENSING AND CONTROL FOR ESTABLISHING AND MAINTAINING DIGITAL IRRIGATION

Akin Cellatoglu[1] and Balasubramanian Karuppanan[2]

[1]Dept. of Comp. Engg., European Univ of Lefke, Turkish Republic of Northern Cyprus Mersin-10, Turkey

acellatoglu@eul.edu.tr

[2]Dept. of EE Engg., European Univ of Lefke, Turkish Republic of Northern Cyprus Mersin-10, Turkey

kbala@eul.edu.tr

***ABSTRACT***

*The remotely sensed data from an unknown location is transmitted in real time through internet and gathered in a PC. The data is collected for a considerable period of time and analyzed in PC as to assess the suitability and fertility of the land for establishing an electronic plantation in that area. The analysis also helps deciding the plantation of appropriate plants in the location identified. The system performing this task with appropriate transducers installed in remote area, the methodologies involved in transmission and data gathering are reported.. The second part of the project deals with data gathering from remote site and issuing control signals to remote appliances in the site; all performed through internet. Therefore, this control scheme is a duplex system monitoring the irrigation activities by collecting data in one direction and issuing commands on the opposite direction. This scheme maintains the digital irrigation systems effectively and efficiently as to utilize the resources optimally for yielding enhanced production. The methodologies involved in extending the two way communication of data are presented.*

## 1. INTRODUCTION

When a land is available in unknown area for plantation and cultivation, deciding to establish a desirable plantation needs a long term investigation of the desirable parameters of the land and its environment. An electronic method of area surveillance and monitoring would be helpful for such tasks. The essential parameters to be sensed for establishing an appropriate plantation unit would be water, wind, temperature and few related items [1, 2]. Water is the main natural resource of irrigation system which keeps the plants to grow efficiently and without which no plantation unit could be established. The continuous availability of water throughout the period of plantation until cultivation has to be ascertained. In order to assess the suitability of the land, long term monitoring such as the period of an year passing through various seasons is needed to be gathered. The second system being irrigation system it performs the monitoring operations during the entire plantation period and based on the data collected necessary electronic commands are issued to the appliances in the site for watering and pest spraying requirements.

## 2. REMOTE SENSING OF ESSENTIAL PARAMETERS FOR PLANTATION

The remote location for plantation is first identified and appropriate transducers for sensing the parameters are installed at chosen points by visiting the site. The transducers are distributed in the identified locations in single or in multiple forms which provide sufficient information for deciding the plants for the area. The transducers generate electrical signals of the concerned parameters under consideration and the signals are conditioned as in conventional manner, digitized to produce binary information and conveyed to the computer through wireless unit.

### 2.1 Parameters to be Sensed

Water is the primary natural resource for the irrigation system for plantation unit and therefore the availability of water should be assessed first and confirmed. There can be natural water reservoirs which would collect water flowing occasionally or continuous flow of water streams arising from rains or from opening of water beds. If there would be hills nearby and if they hold snow during winter they also supply water in summer to the reservoir due to melting. By physical inspection about the deepness of the reservoir and the capacity of the storage in long term the water level accumulated in the reservoir during different days of the year has to sensed and monitored. There are several water level sensors used in practice [3-5]. We use here the capacitive type of transducer where the capacitance changes due to change in dielectric constant resulted due to the level of the water. The capacitance is converted into frequency [6] and frequency in turn is converted into voltage by a frequency discriminator and finally voltage is developed proportional to water level. During irrigation periods of the plantation unit the moisture contents of the soil has to reach the desired level for improved growth of the plants. Also, assessing the natural moisture contents in the absence of plants would provide information about the selection of plants for that location. A similar capacitive type of pickup to get a voltage signal proportional to moisture content of the soil is used for this purpose.

The daily and seasonal variation of ambient temperature is an important factor to be monitored throughout the period of investigation. This would also help in deciding the type of plants for the plantation. We use Thermistor type of transducer included in Wheatstone bridge to get a voltage signal proportional to temperature. Appropriate linearization [7] is performed by hardware at the transducer level itself. Natural wind flow within limits helps to make the growth of the plants. Heavy wind will have ill effect on the growth of certain type of plants. Therefore the velocity of wind flow is another desirable parameter needed to be sensed. In practice, several types of wind flow sensors used are of ultrasonic type. As this type of sensor cannot align itself to the direction of wind we use miniature windmill [8] to sense the wind flow efficiently. Therefore, despite change in direction the wind flow it produces maximised output. Ambient light available due to solar radiation helps the growth of most plants. This is another desirable parameter to be sensed. For this purpose, standard photo sensor circuit is used as ambient light transducer.

Humidity of the environment influences the growth of most plants. It represents amount of water vapor present in air. Standard humidity sensor, hygrometer, is employed to monitor the humidity [9]. Fire detectors may help in assessing how best the area is immune to catching fire with plantation of certain type of plants. Standard fire detectors [10] are used in this project and kept at selected locations evenly distributed in the field. With this the frequency of catching fire in summer and also in winter can be monitored. For assessing the consistency of the fertility of the soil its pH value is monitored [11] throughout the year during different conditions of climate. As manual inspection of the site is only rarely done, the movements of forest animals over the area also are monitored. Although sensors could be arranged for sensing moving objects for such requirements, it is desirable to capture the video image of the site and transmitted to the remote location. Frozen still images transmitted would help in assessing several other aspects of the site. Heavy fire, animals, hunters and group of birds or grasshoppers which would be bothering the crops are easily captured by video and visualized at the remote end.

If there is a water stream or water channel connecting the site to outside township or village, availability of water flow is sensed by installing standard flow sensors at selected points in the stream. If water is available throughout the year the crops cultivated could be transported to external places through boating service. These are the desirable parameters to be sensed. If additional parameters are needed to be studied they can easily be included in the system for sensing and transmission. Appropriate reservations are made in hardware and software for insertion of the extended parameters in future.

## 2.2 System Design for Remote Sensing and Communication

The sensors are located at chosen locations of the area identified for plantation. The signals from the transducers are conveyed to a computer installed in base station setup in the plantation site. The computer in turn arranges to upload the data to website. Once it is uploaded to website it can be monitored in a computer at any remote location and an analysis can be made. In the past, digital control of a model digital factory performed through web [12] was reported. The features of the project are suitably modified and adapted to the present application yielding enhancement in performance.

Figure 1 shows simplified schematic of the system monitoring remotely the parameters. In the base station of the plantation site a PC, lab top or notebook is fixed with wireless card attachment accessing the server and the internet. Since the transducers are located at different places, use of cables connecting to computer may not be feasible. Therefore, modular wireless communicating unit such as the popular USB data card is used for computer providing access to internet.

All transducers have mini wireless communicators extended to them through which the data are continuously transmitted to the base station. The data from all wireless transducers are collected in the receiver extended to PC and saved in memory as a table which is updating its contents continuously. The computer being extended to internet the table is updated periodically in the website. In the remote PC after giving the user name and password the website is open and the table can be visualized. For use in analysis package, the table is periodically transferred and saved in computer hard disc. The power output of the wireless unit employed depends on the range between the transmitter kept in site and the receiver extended to PC in site.

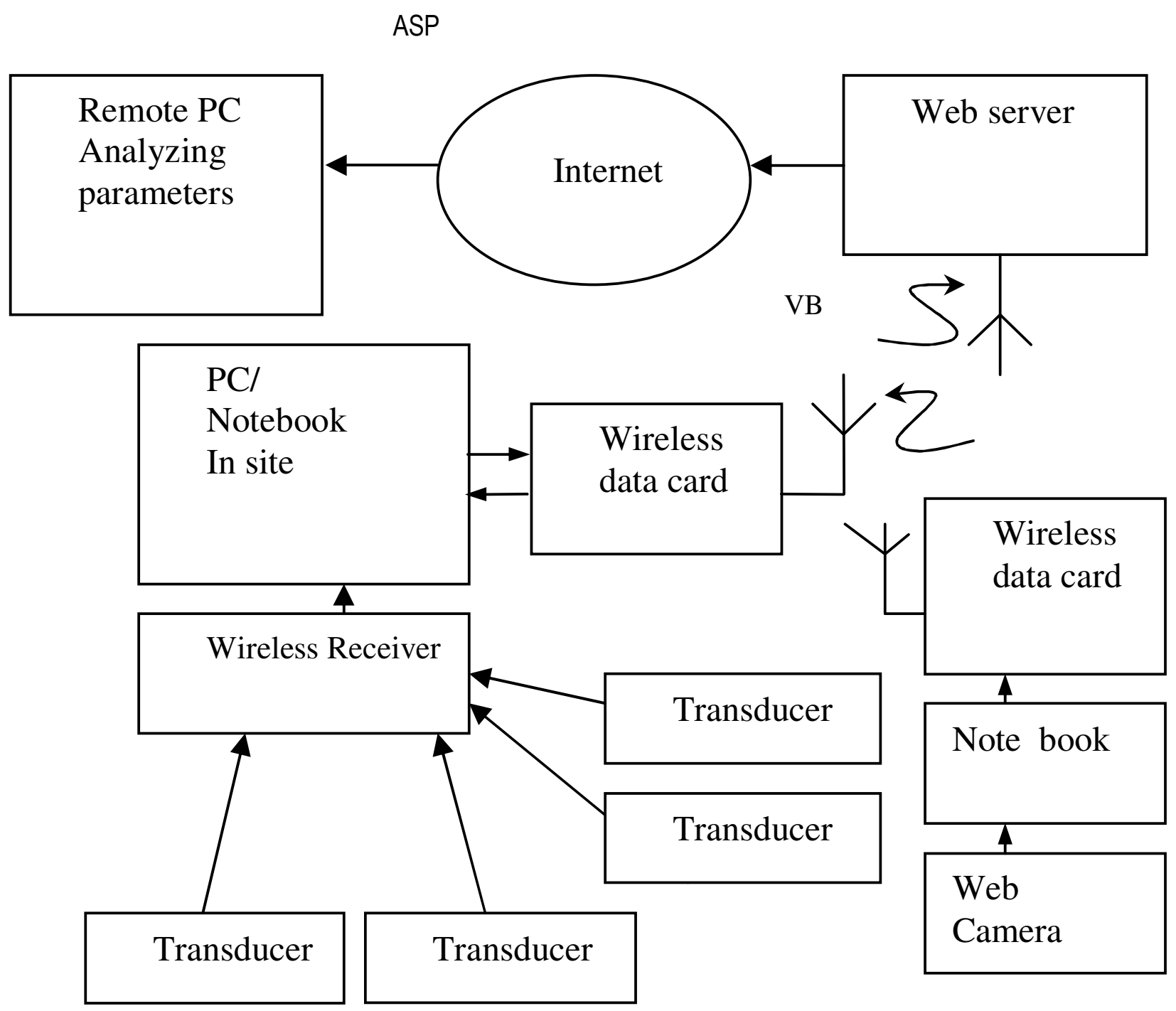

Figure 1. Remote sensing of plantation parameters

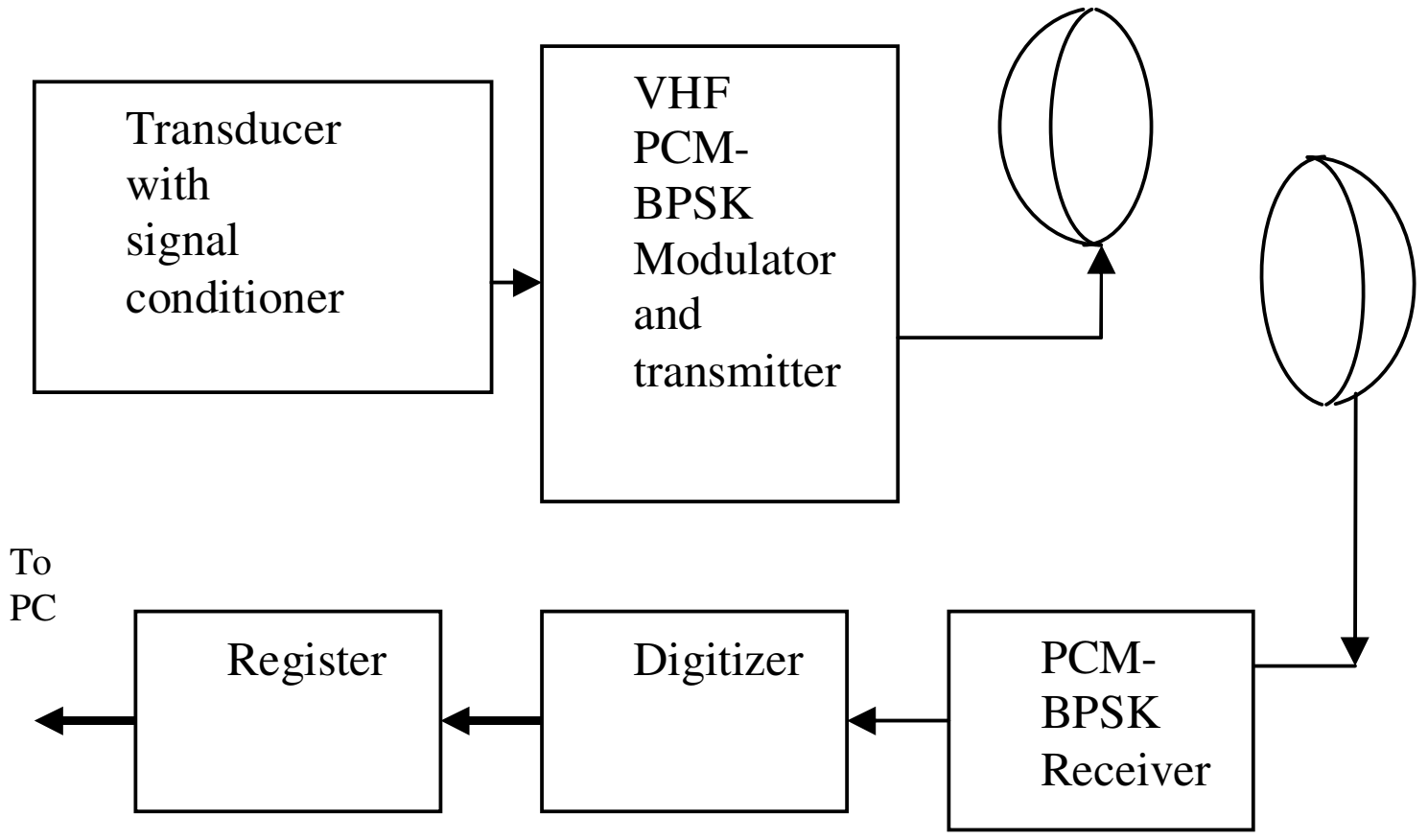

Figure 2. Wireless communication of the sensed signals

Figure 2 shows the simplified schematic of the wireless transducer data communication system. The transducer analogue signal is digitized with a modular flash ADC (Analogue to Digital Converter) [13] and transmitted as PCM-BPSK signal as to ensure higher probability of detection. All transducer signals are transmitted via the wireless communicating unit except the transmission of video image got through web camera. Video image of web camera is saved in a notebook and directly uploaded to internet through wireless data card. If power supply is not available in the site of exploration then modular power supply unit [14] generating electric power from mini solar panel and mini-windmill could be included in each transducer sensing and communicating the signal.

### 2.2.1 Software Medium and Link to Various Communication Units

The link to internet from webmail server is made through Visual Basic. As internet is accessible for the system manager as a client stationed at remote location he can view the table as provided by ASP. The server running the VB program gets the data sent from the PC and with the help of ASP put these data into webpage for accessible to PC at farther end.

### 2.2.2 Monitoring the Conveyed Parameters in Remote PC

The Plantation website is open at the PC. In order to open it user name and password are needed. When the website is open a widow displaying the sensed parameters as a table in the screen of the monitor. Table 1 shows its form of appearance wherein various sensors employed are listed and the present data received from source are displayed. Columns 3 and 4 are meant for the tests performed on various transducers at the site. Tests are done periodically as set with the transducers by extending the standby unit to the wireless transmitter during the testing period. The column 3 indicates the time before which the test was performed on the concerned transducer. The status of the test is indicated in column 4. If the test status shows error, then the standby transducer continues to work with the wireless unit.

The data in numeric form has relevant units incorporated on them. For instance the temperature showing as 25.5 is representing it in degree Celsius. The pH value between 4 and 7 shows acidity which is required for most plants.

The current data appearing in the table with the test information are saved in reserved memory locations as required for analysis package. From this current data received a set of relevant data are extracted and gathered in memory on FIFO basis. If a transducer has shown error status and later the standby transducer replacing this also shows the error status then the transducer needs replacement. Based on the status of the test received manual visits need be made to the site for replacing them with new ones.

Table 1. Status window appearing in website

| **Sensors** | **Present data** | **Test done for sensor- before seconds** | **Test status** |
|---|---|---|---|
| Temperature | 25.5 | 30 | OK |
| Water level in lake | 40 | 30 | OK |
| Water level in overhead tank | 2.5 | 30 | Error |
| Wind flow | 20 | 30 | OK |
| Moisture contents of the soil | 0.5 | 30 | OK |
| pH value of the soil | 5.0 | 30 | OK |
| Humidity | 0.6 | 30 | OK |
| Fire and smoke | 0.2 | 30 | OK |
| Water flow in stream | 0.7 | 30 | OK |
| Light sensor | 0.3 | 30 | OK |

## 2.3 Wireless Sensor Node for Transmitting the Data

As an alternative to internet for communicating the data wireless sensor network can also be used. The wireless sensor node is a part of wireless sensor network. The node known as sink node attached with the computer with USB port collects information and transmits to other nodes. This node can be extended to satellite link with the use of appropriate interfaces. In the recent past, by linking the wireless sensor network to a satellite transmission, the data from a set of triangulated radars installed for regional exploration has been reported [15]. This approach is now extended to the present application for transferring transducer parameters to remote monitoring PC.

Eight channels are normally extended to the node with USB connection and if more channels are needed to be included they have to be connected in time multiplexed form. In this application of remote sensing and communication, as there are ten transducers, they are extended to the node with appropriate multiplexing.

The simplified block diagram of the wireless sensor node is shown in Figure 3. A microcontroller installed is the heart of this node. The microcontroller reads all sensors data and saves them in memory at reserved locations. Through appropriate USART (Universal Synchronous Asynchronous Receiver Transmitter) interface the data are transferred to the STX-2 [16] Satellite Transmission board. It radiates the signal at 2.4GHz. The satellite network receiver at the other end gets these data and puts them in the monitoring PC performing as node in wireless sensor network.

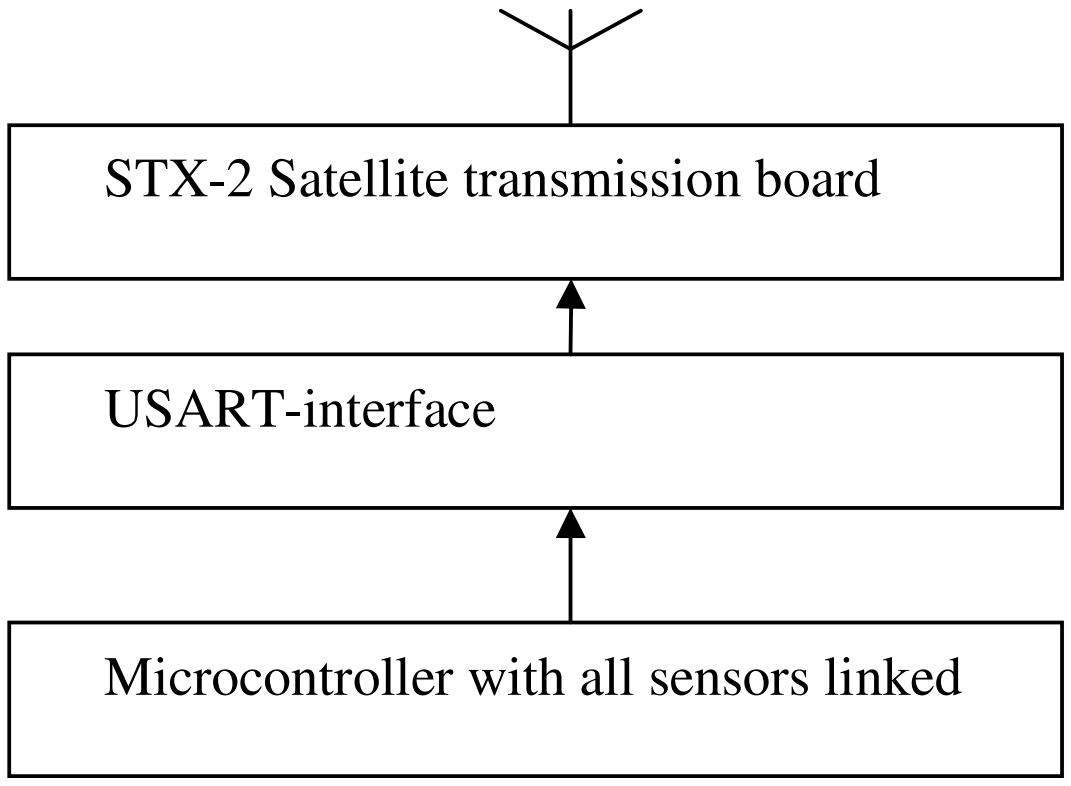

Figure 3. Simplified block diagram of satellite based wireless sensor node

#### 2.3.1 Comparison of Wireless Sensor Network and Internet for Transmitting Transducer data

The wireless sensor network is being used for different applications without access to satellite link. Each node can transmit and receive information as to satisfy the environmental requirements. The processing of information is done at the sink node linked to a computer. It has however restricted radio range.

In this project as the remote monitoring PC is situated at a far place, the help of satellite link was necessitated in transferring the transducer parameters. The hardware involvement in using the internet communication and the satellite link are more or less the same. The cost involvement in using internet and satellite link has to be explored depending upon the involvement of number of channels. The time delay of information reaching the PC is not a critical issue in this application and therefore the performances of both mediums are well acceptable.

## 3. DIGITAL IRRIGATION SYSTEM

In the remote sensing and transmitting system described above the parameters sensed at one place are monitored in computer in a remote location and there is no control activity involved in it. The objective of that implementation was to explore the possible agricultural parameters of the land for its suitability of establishing an appropriate plantation unit. This is therefore a simplex system transmitting data in one direction from the site to remote PC.

The remote control system employed in digital irrigation system is duplex system where sensed parameters flow in one direction and control signals flow in the opposite direction. This scheme is applied to a remotely located irrigation system where the plantation already exists. We use partly the facilities available in the simplex remote sensing system and incorporate remote control activities so as to fulfil the requirements of establishing duplex system. Therefore, we observe in a computer the sensed parameters arising from the irrigation site transferred through internet. After making an analysis of the sensed parameters control commands are issued through the same internet as to switch ON or OFF selected appliances in the irrigation system.

Regular routine works such as opening and closing the valves at a predetermined periodicity are performed on the site. The control signals arriving from remote PC overruns this activity as to keep the irrigation efficiently.

### 3.1 Parameters Sensed for Irrigation

As the main objective of irrigation system is to maintain required moisture contents in the soil for entire duration of cropping, the determination of the moisture contents becomes essential parameter for this task. The desirable parameters to be sensed in this case are as follows.

*i.* moisture contents of the soil
*ii.* temperature
*iii.* wind flow
*iv.* light
*v.* water level in the reservoir lake
*vi.* water level in the overhead tank.

Provisions have been given for including additional parameters in the sensor table.

## 3.2 Irrigation System Design

Figure 4 shows the simplified block diagram of the digital irrigation system maintained through internet. The essential parameters sensed by transducers at various locations are transmitted to the computer at irrigation location and then uploaded in the website, all done similar to the previous simplex remote sensing unit. The computer at the irrigation spot is interfaced to a PIC microcontroller which takes care of

*i.* controlling the devices as per the command issued by remote PC through website and

*ii.* controlling the regular routine irrigation activities as per its program set already.

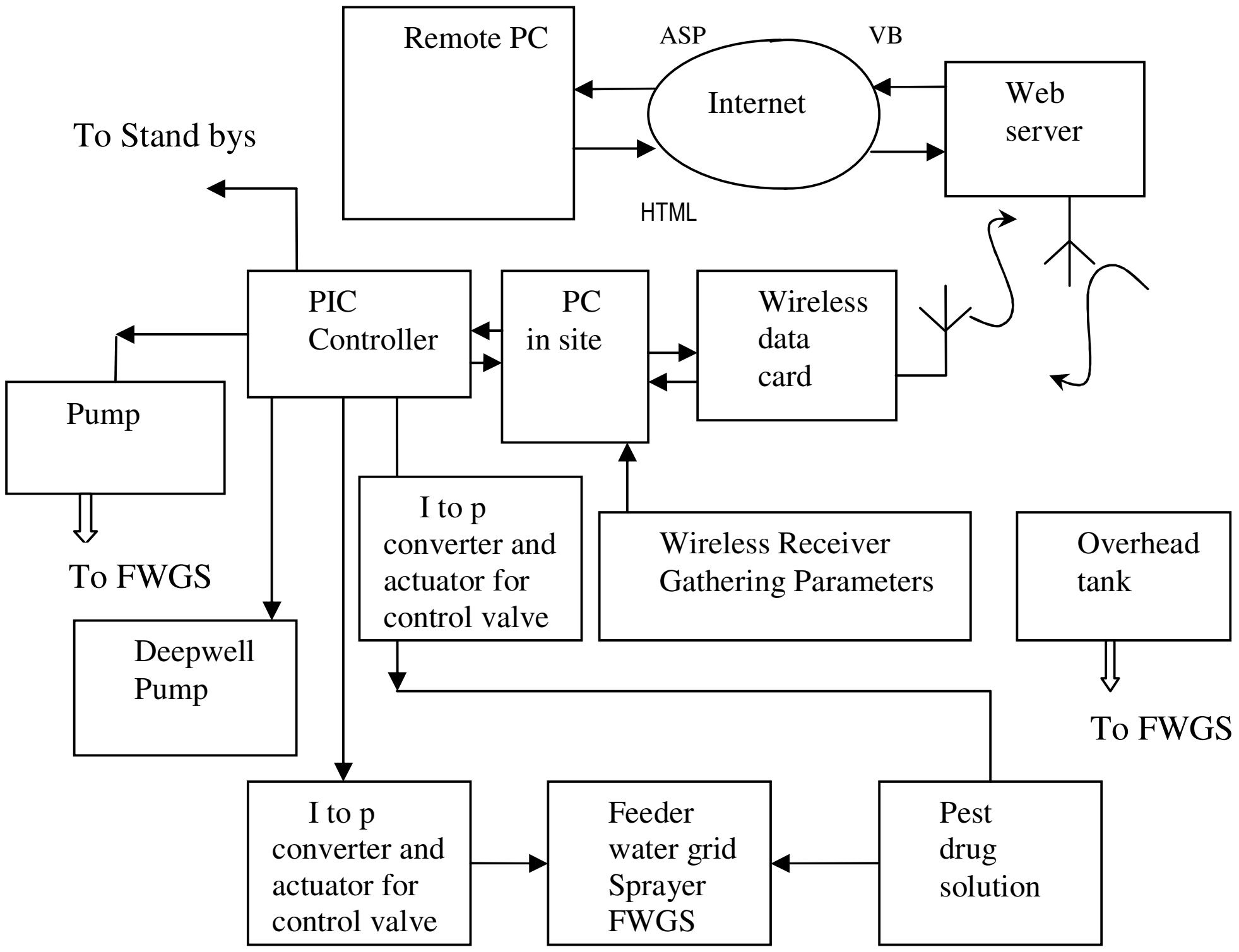

Figure 4. Remote control of digital irrigation system

As described before, there are two water pumps included in the system. The deep well pump is used to pump the underground water from rigged well to the overhead tank supplying water for irrigation. Natural water reservoir like lake is another source supplying water to the overhead tank. In order for the irrigation to perform smoothly the water level in the overhead tank should be maintained at specified level. The water from overhead tank or pumped water from reservoir can be connected to the feeder pipe by performing electronic control.

When the water level of the reservoir is good enough then that the pump connecting reservoir to feed water grid sprayer (FWGS) is preferred to be switched ON first. If it is not enough then the deep well pump of rigged well is operated to fill the overhead tank. The overhead tank has required head to feed the FWGS. Decisions are made to switch ON a particular pump depending upon the moisture contents, water levels and past history of the switching made on the pumps.
The FWGS can supply overhead tank or pumped water from reservoir as executed from remote PC as well. As a consequence of command received from PC the PIC issues signals to operate the control valves installed with the feed water sprayer. The electrical current from PIC is converted into pneumatic pressure actuating the respective control valve. The pest drug solution can be mixed with water and sprayed to the plants. Pest spraying is performed at required periodicity.

When the website is open in the remote PC two windows can be seen in the screen. The first window is like a table shown in Table 1 informing the status of the transducer parameters included in the system. The second window is meant for issuing control action and it appears as shown in Table 2. Only two pumps are included in the system for control and the table shows their present status. Also the present status of irrigation feed whether connected to overhead tank or cut OFF is also indicated. There is another column marked as control column which can be changed by the manager at the remote PC. For instance, in the table the present status of deep well is shown as OFF and the control also is shown as OFF. After making an analysis and decision if the deep well pump has to be switched ON the manager just types 'ON' in this field. This issues necessary command in the PIC at the site to switch ON the deep well pump. In the case of switching the standby transducer, the user clicks the present status/transducer and uses the ↓ key to see the status of transducers one by one. For instance, overhead tank water sensor needs the standby to be connected then he selects that sensor and also selects the connect-standby option to send this command to the PIC at the other end.

Table 2. Control table window appearing in website

| **Control Devices** | **Present Status/ Transducer** | **Control Command** | **Duration of Execution** |
|---|---|---|---|
| Deep well pump | OFF | OFF | NA |
| Pump from lake | OFF | ON | 30 |
| FWGS from pump or lake | OFF | ON | 100 |
| FWGS from Drug Solution | OFF | OFF | NA |
| Standby Transducer/ Select | Over head water sensor | Connect Standby | 1000 |

The web server running the VB program checks for any command issued from the remote manager PC. If a command exists then it formulates the appropriate bytes and sends them to the PC at the irrigation spot for issuing necessary control actions to PIC for switching on/off of the pumps, control valves or the standby transducer.

### 3.2.1 Microcontroller-PIC Hardware Control

The PIC microcontroller connects the water tap daily to the feed tubes going to all plants for a prefixed time as set in the program. The time period is changed by the remote PC after making the analysis of moisture contents of the soil, temperature, water levels in the reservoir and the overhead tank and the ambient light. Therefore, in the PIC monitoring program it checks for the commands issued from the server in response to remote Manager PC. Therefore, while performing the regular daily routines it also actuates the commands issued from the remote PC. The main control actions arising from remote end are switching ON or OFF the two pumps and the irrigation feed tube.

Circuit diagram to switch ON or OFF the appliances such as pump is shown in Figure 5. The PIC processor issues just one bit (high) as to switch ON or (low) OFF the motor-pump as to pump the water accordingly. The opto-coupler provides necessary isolation for transients and EM interferences.

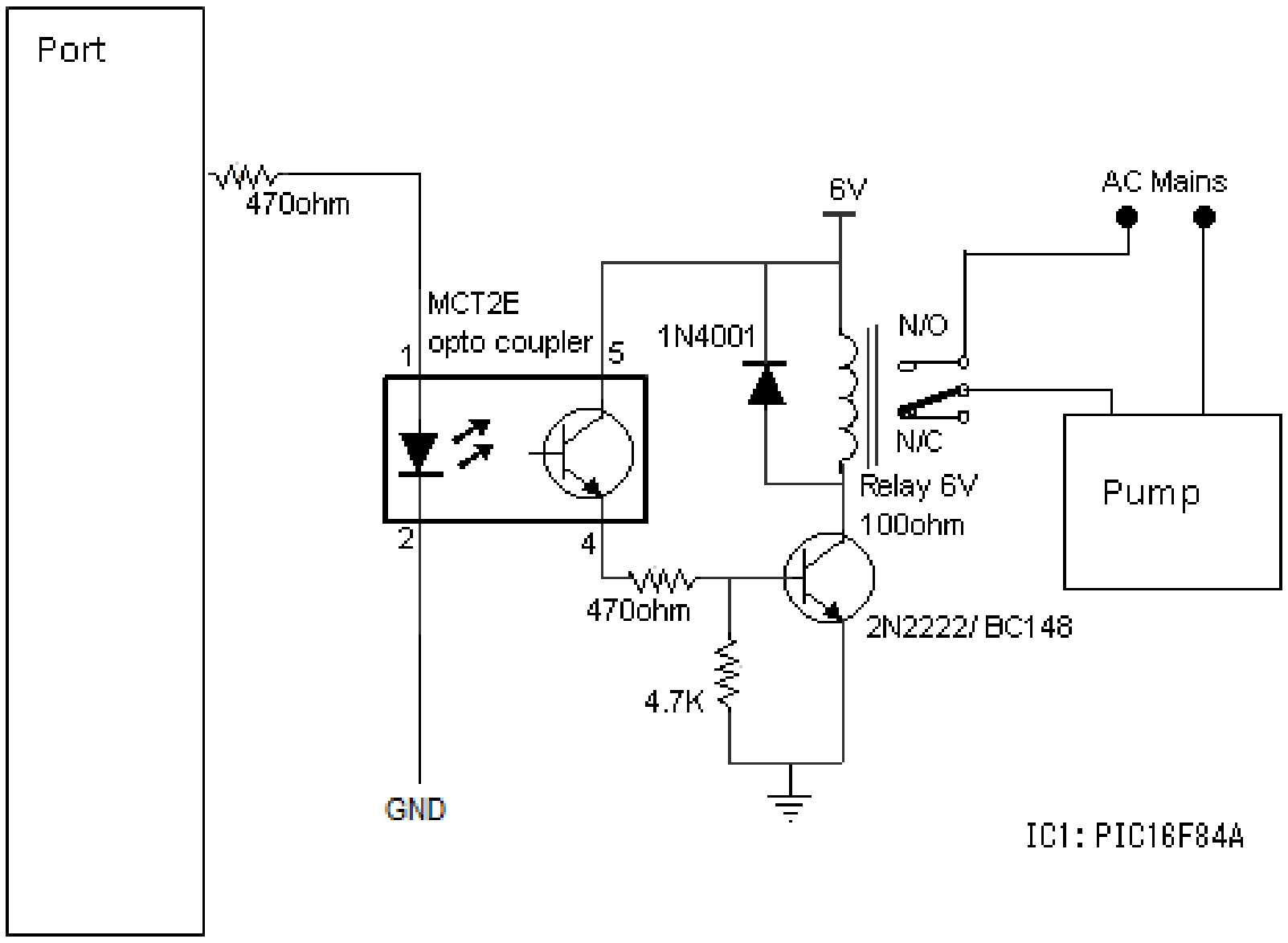

Figure 5. PIC circuit for controlling a relay

## 4. SYSTEM ANALYSIS

A performance analysis about the remote sensing and control activities performed in digital irrigation unit is made and presented here. Also performed is the financial analysis of the irrigation system producing cumulative cash flow.

### 4.1 Financial Analysis with Computation of Expenditure Involved

In order to assess the financial involvement and benefits of the remote control employed for irrigation two methods of analysis are performed. The performance of the remote control techniques over the manual operations are assessed based on the available data and presented. The comparison is made between the manually controlled irrigation and digitally controlled irrigation. Even in manual control method, transducers are employed to get the information about the soil and others.

The initial investment is almost same for the digital irrigation and the manual control used for irrigation. Period of growing and cultivation of crops is different for different kinds of plants. Although some kinds of plants yield production shortly within a month, most plants need a period of three months to six months. Therefore we presume the growing and cultivation period of six months for the present digital irrigation scheme performed by remote control. Figure 6 shows the financial involvement of both schemes for a period of 6 months. The expenditure shown is the normalized expenditure made for the area of coverage of the plantation of 10 hectares. Although the initial expenditure involved in both the schemes are almost same the digital irrigation yields less running expenses compared to manual controlled irrigation.

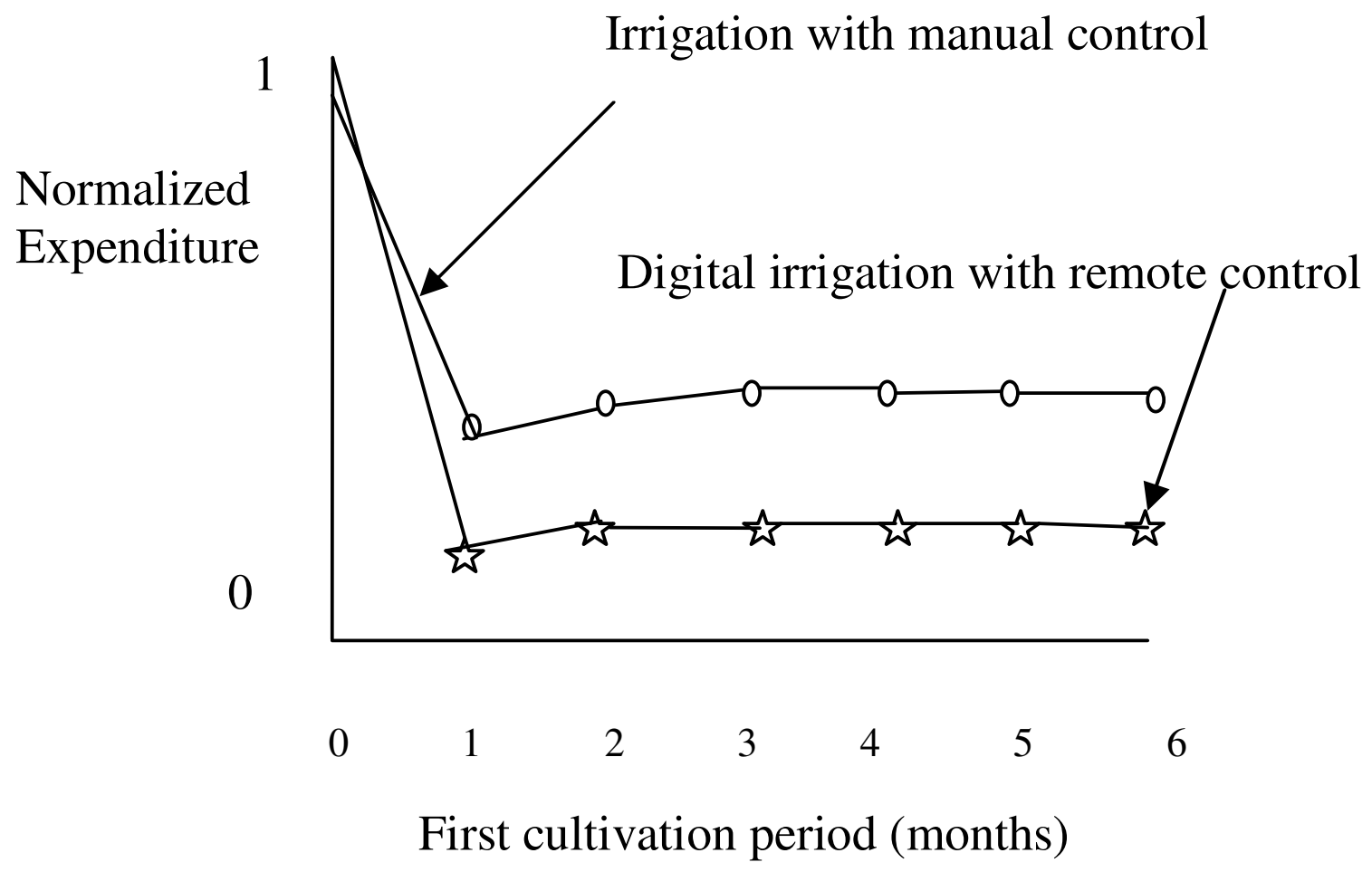

Figure 6. Financial involvement during first cultivation period

## 4.2 Analysis of Irrigation Digital Irrigation with modified Retscreen software

The Retscreen analysis is a financial analysis software which can produce the cumulative cash flow of the renewable energy system when the required data are given as input to the software [17, 18]. We now emulate the Retscreen software with suitable modifications made to it such that the financial analysis of digital irrigation system involving remote sensing and control. Figure 7 shows the tasks performed by modified Retscreen analysis software package.

To the settings of the package inputs are given describing the project type and technology involvement and also the climate factors influencing essentially the soil conditions, moisture and wind flow. The settings and environment are applied to irrigation model already installed in the package. The model covers various activities involved in the remote control of digital irrigation.

The input of the model are financial involvements involved and the initial cost spent for the transducers, local communicating units, ground station PC and the initial cost spent for irrigation unit posted in internet. The outcome of the irrigation model are given to the financial analysis package which in turn produces the annual savings compared to the manual control scheme employed for irrigation and the cumulative cash flow.

The cumulative cash flow for the possible life span of 10 years is produced by the modified Retscreen software and shown in Figure 8. It shows that the amount invested for the digital irrigation can be recovered in first 2 years and after which it yields profit to reach nearly 7 times that of the investment in remaining 8 years.

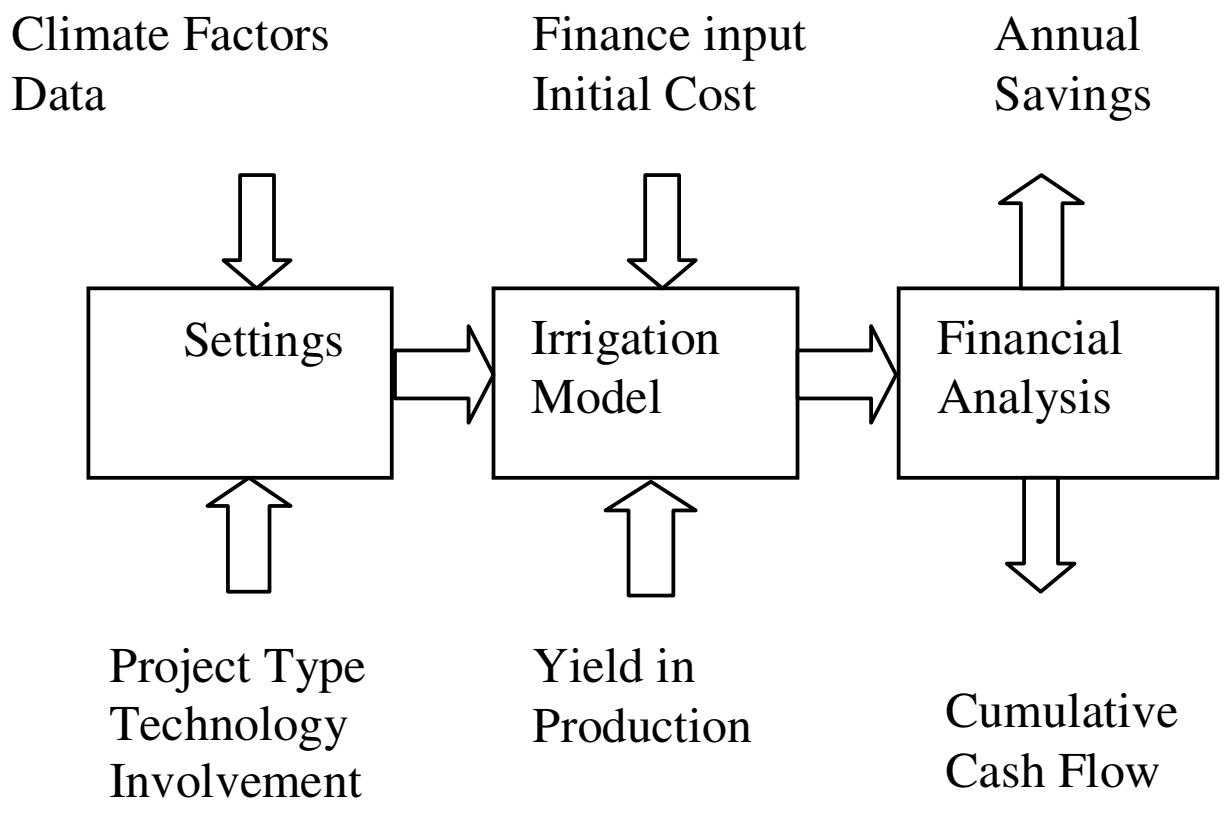

Figure 7. Block schematic of the modified Retscreen software

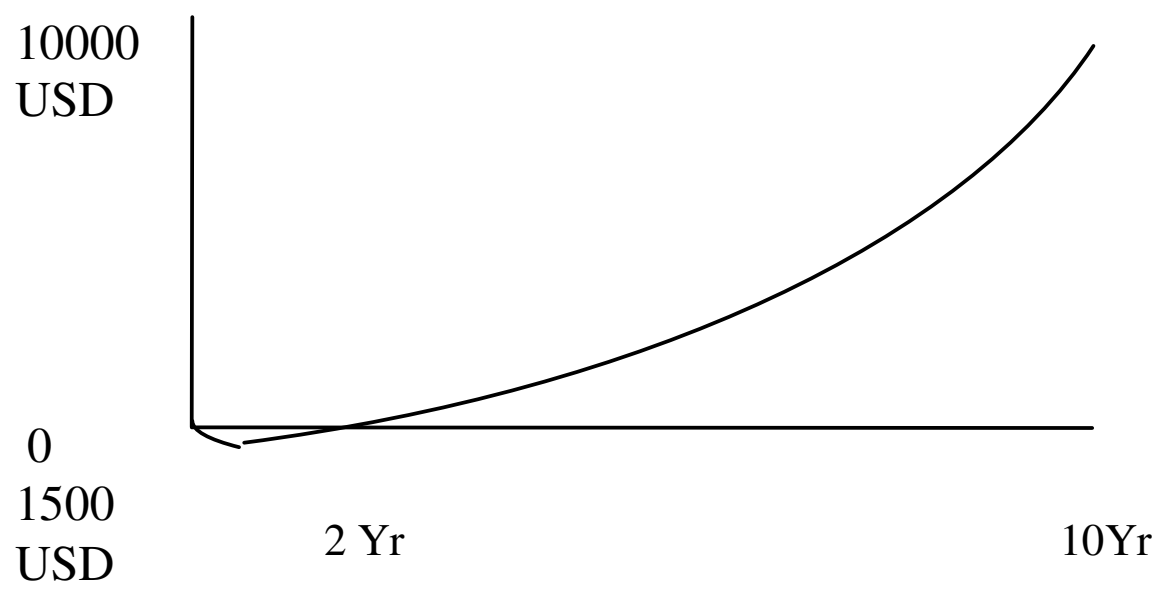

Figure 8. Cumulative cash flow indicating savings compared to manual control scheme

## 4.3 Performance Analysis of the Remote Control of Digital Irrigation

The systems described here employ internet as the tool for communication. Security concerns, testability, maintainability, limitations and extendibility are the major factors to be analyzed as a means of assessing the performance of the systems.

### 4.3.1. Security Concerns

By giving user name and password the website in the remote Manager PC cab is open at any time. As soon as the website is open one can see the status window and control window. As the website demands user name and password it ensures security and no unauthorized person could issue wrong control. By making the password issue a complex concern the security aspects could be strengthened.

### 4.3.2. Testability

The system issues test signals and their test status are indicated. Furthermore, as to ensure the accuracy semi automatic testing is performed by visiting the plantation site and checking the parameters with various sensors and the magnitudes of signals can be verified in the website. Test signals at different levels are inserted and watched in the website for their changes and also the time constant involved are tested. The time constant depends on the communication path between the source and the destination and the bit rate of transmission. The status of sensors and control appliances are checked with other available communication mediums as well. By these cross checks of the data reliability of the remote sensing system and remote control schemes are ascertained.

### 4.3.3 Maintainability

Malfunctioning of any of the devices can provide wrong information about the parameters and the analysis package might produce an error. This is avoided by properly maintaining the devices in working conditions. Each transducer is backed up with a standby unit and a local simple testing process is carried periodically as to switch ON the standby unit as active transducer and test the output of both of them. If the outputs do not tally, the error signal is communicated along with the transducer signal to the website for taking corrective measures at the other end. The transducers employed are fool proof ones that it provides self generated error output, if any.

### 4.3.4 Limitations

The sporadic interruptions occurring randomly in the internet such as disturbances in the server level or in the communication network influence the remote data acquisition system. Nevertheless, as the decision is not taken instantly such disturbances do not have any impact on the system performance.

### 4.3.5 Extendibility and inclusion of several other parameters

This system has no limitations in extending additional parameters and control appliances needed for different environment in digital plantation or digital irrigation system. When hardware appliances are brought under remote control it needs revision in the software also. This is only a minor task and could easily be performed.

## 5. Discussions and Conclusion

The system performs the remote sensing and remote control activities without the manual observation and attention in the site. Being an automatic system unavailing manpower and providing information for long period like a year it is highly preferred in several instants as to decide the type of the plantation unit in the identified remote site. A watch span of one year undergoes all seasons encountered and provides sufficient information for the judgement.
As internet usage has become a common means of sharing and exchanging information between users its usage for implementing remote control of a digital irrigation system simplifies its communication process without calling for special communication systems. On the other hand internet is still growing sector with more and more incorporation of additional strategies and overheads and this act would certainly enhance the features of internet based remote sensing and remote control systems.

In place of the single channel wireless data transmitter installed at each transducer for transmitting information to PC at the base station, one can think to use wireless sensor node and network. In this case it has the advantages of reliability ensured by the wireless sensor network. On the other hand this activity uses only a part of facilities available in the wireless sensor node and utility part would be low.

Only few control actions are included in the irrigation system reported. It would be easily possible to extend further with as many control actions as needed for the plantation site. The system accepts the desired number of sensors and control channels with appropriate changes made in hardware unit and software package.

The video channel conveyed to the remote PC provides information about the accidents, fire hazards, movement of animals and many other things. Selected frames of images can be frozen and saved in memory for future analysis purposes.

The remote control techniques employed here help maintaining the irrigation system which is far away from the control place. The water resources are economically utilized after testing the conditions of the soil. If needed, alarms can be arranged to appear in the remote station monitor drawing the attention of the manager when any error occurs.

## ACKNOWLEDGEMENT

The authors would like to thank the Rector of European University of Lefke for providing financial support for this project.

## Authors

Akin Cellatoglu received his Bachelor's degree in Electronics and Communication Engineering from Eastern Mediterranean University, Turkish Republic of Northern Cyprus in 1996. He obtained his M.Sc degree and Ph.D degree from the University of Surrey in 1998 and 2003 respectively. Dr.Akin is with the Computer Engineering department of European University of Lefke, Turkish Republic of Northern Cyprus since September 2003. His fields of specialization are in video codec systems, multimedia and communication networks

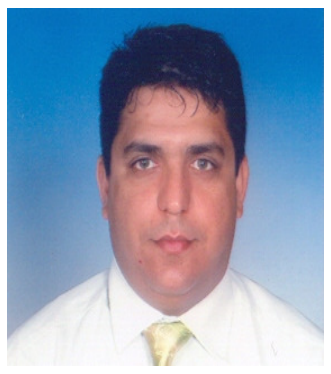

Balasubramanian Karuppanan received his Bachelor's degree in Electronics and Communication Engineering from PSG College of Technology, Madras University, India in 1971. He obtained his M.Tech and Ph.D degrees from the Indian Institute Technology, Madras in 1976 and 1984 respectively. He was working in Calicut University, India, during the periods 1972-1990 and 1995-1998 in various positions as Lecturer, Asst. Professor and Professor. In 1988, he did post doctoral research at Tennessee Technological University, Cookeville, TN, USA under Fulbright Indo-American Fellowship program. He joined Cukurova University, Adana by June 1990 as Professor and worked there until Feb1995. By 1996, he was granted the Best College Teacher's Award from the University of Calicut, India. From Feb1998 onwards, he is with the Faculty of Architecture and Engineering of European University of Lefke, Turkish Republic of Northern Cyprus. Dr.Balasubramanian is a life member of Instrument Society of India and the Indian Society for Technical Education. His fields of specialization are in 3D imaging and microprocessor based systems.

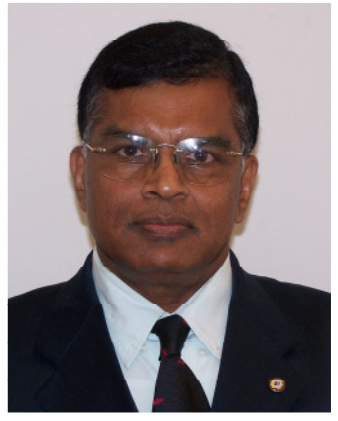